\documentclass[amssymb,prd,aps,showpacs,nofootinbib,preprint]{revtex4-1}
\usepackage{latexsym}
\usepackage{graphicx,color}
\usepackage{graphicx}
\usepackage{amsmath,amssymb}
\usepackage{amsmath}
\usepackage[T1]{fontenc}
\usepackage[hang,small,up]{caption}
\newcommand{\sgn}{\mathop{\mathrm{sgn}}}
\makeindex
\begin{document}
\title{Critical behavior of noise-induced phase synchronization}
\author{Pedro D. Pinto$^{1}$}\author{Fernando A.
Oliveira$^{2,3}$}\email{fao@fis.unb.br}\author{Andr\'e L.A. Penna$^{2,3}$}
\affiliation{Universidade Federal do Oeste da Bahia, CP 47850-000, BA, Brazil,$^{1}$}
\affiliation{Instituto de F\'{i}sica, Universidade de Bras\'{i}lia, Brazil$^{2}$\\
International Center for Condensed Matter Physics\\CP 04455,
70919-970 Bras\'{i}lia DF, Brazil,$^{3}$}


\begin{abstract}

In this article, we present a systematic study of the critical behavior of phase oscillators with multiplicative noise from a thermodynamic equilibrium approach. We have already presented the thermodynamics of phase noise oscillators and mapped out in detail the behavior of free energy, entropy, and specific heat in a previous work [P. D. Pinto, F.A. Oliveira, A.L.A. Penna, Phys. Rev. E 93, 052220 (2016)], in which we also introduced the concept of synchronization field. This proved to be important in order to understand the effect of multiplicative noise in the synchronization of the system. Using this approach, our aim is now to study analytically the critical behavior of this system from which we derive a fluctuation-dissipation relation as well as the critical exponents associated with the order parameter, specific heat, and susceptibility. We show that the exponents obey the Rushbrooke and Widom scaling laws.

\end{abstract}
\date{\today }
\maketitle

\section{Introduction}

It is recognized that one of the major conceptual advances in the understanding of synchronization phenomena in physical systems, was its interpretation as a phase transition phenomenon by Yoshiki Kuramoto. In fact, making analogies with the mean-field theories for magnetic systems, Kuramoto developed its approach by defining an order parameter for your coupled oscillators model, which allowed the characterization, under various conditions, of the transition of the oscillator system for the synchronized state \cite{Kuramoto1}. It was subsequently found that the distribution of natural frequencies influences the type of phase transition model, which can exhibit continuous or discontinuous transitions and therefore different critical exponents \cite{Basnarkov1,Basnarkov2}. Moreover, a phase transition has been also characterized by a symmetry breaking and a discontinuity in the first derivative of the order parameter in coupled oscillators \cite{Ewa}. Other studies have shown that in addition to the frequency distribution, the form of the coupling function can also change the critical exponent of the order parameter in the transition \cite{Daido1,Crawford}.

Although much has been discussed about the critical exponent of the order parameter, there is still a lack of studies that systematically address other critical exponents that characterize the complete theory of phase transitions of these coupled phase oscillators systems. In part, this is due to the lack of proposals that deal with the thermodynamic extension of these models, especially the absence in the literature of the field concept associated with synchronization, as well as an analytical expression for susceptibility. Recently, several studies have been reported \cite{Daido2,Yoon,Hong} that address the susceptibility to phase oscillators but give priority to its numerical analysis.

In this article we present a broad analytical approach to the critical behavior of phase oscillators, where we calculate the main mean-field critical exponents for the generalized Kuramoto model systematically developed from a thermodynamic equilibrium approach \cite{Pinto}. We analyze the critical behavior of the various thermodynamic quantities involved, such as entropy, free energy, specific heat, synchronization field and susceptibility. We derive a fluctuation-dissipation relation which connects the order parameter fluctuation with susceptibility. Moreover, we show that the obtained exponents corroborate the well-known scale relationship of Rushbrooke and Widom.

\section{The model}

In previous work \cite{Pinto}, we introduced the It\^{o} stochastic differential equation for phase oscillators in the form
\begin{equation}
\label{eqq}
\dot{\theta_i}=\omega_{i}+f_{i}(\{\theta\})+\sqrt{g_{i}(\{\theta\})}\xi_i(t)\,,
\end{equation}
where $\omega_{i}$ are the natural frequencies of the oscillators, and the drift force $f_{i}(\{\theta\})$  and the noise strength $g_{i}(\{\theta\})$ are general functions of the phases $\{\theta\}=\theta_{1},...,\theta_{N}$ and $\xi_{i}$ is a Gaussian white noise which obeys the relations
\begin{equation}
\label{fdt}
\langle\xi_i(t)\xi_j(t')\rangle=2D\delta_{ij}\delta(t-t')\quad\mbox{with}\quad\langle\xi_i(t)\rangle=0\,,
\end{equation}
in which $D$ is the diffusion constant.

We have already demonstrated that the equation for identical coupled phase oscillators, whose dynamics is governed by Eq. (\ref{eqq}), can be reduced to the mean-field approach as
\begin{equation}
\dot{\theta_i}= f(\theta_i)+\sqrt{g(\theta_i)}\xi_i(t).
\end{equation}
For this case, the functions $f_{i}(\{\theta\})$ and $g_{i}(\{\theta\})$ take the forms
\begin{eqnarray}
        &&f_{i}(\{\theta\})=f(\theta_i)=rK\sin(\psi -\theta_i)\\
        &&g_{i}(\{\theta\})=g(\theta_i)=1+r\sigma\cos(\psi - \theta_i)\,,
\end{eqnarray}
where the drift term is controlled by the coupling constant $K$ and the noise strength is driven by the noise coupling $\sigma$ that determines the intensity of global modulation of the noise. The order parameter $r$ of the system as well as its average phase $\psi$ are defined by
\begin{equation}
re^{i\psi}=\frac{1}{N}\sum^{N}_{j=1}e^{i\theta_{j}}\,\,.
\end{equation}
Here, $r$ measures the phase coherence, {\it i.e.}, for $r=1$, the system is fully synchronized, whereas for $r=0$, the system is fully incoherent. A partially synchronized state is obtained when $0< r < 1$.

When the interactions of the oscillator $\theta_i$ with the other oscillators of the system are no longer considered individually, but in terms of the mean-field effects of the system acting on the oscillator $i$, we can omit the index $i$ of the individual oscillator $\theta_i=\theta$, such that Eq. (3) is given by
\begin{equation}
\dot{\theta}= f(\theta)+\sqrt{g(\theta)}\xi(t).
\label{eq:lang_mult}
\end{equation}

The corresponding Fokker-Planck equation from Eq. (\ref{eq:lang_mult}), in It\^{o} prescription, is given by
\begin{eqnarray}
\frac{\partial\rho}{\partial t} &=& D\frac{\partial^2}{\partial\theta^2}[g(\theta)\rho]-\frac{\partial}{\partial\theta}[f(\theta)\rho]\nonumber\\
\nonumber\\
&=&D{\partial^2\over\partial\theta^2}\big[(1+r\sigma\cos(\psi-\theta))\rho\big]-{\partial\over\partial\theta}\big[rK\sin(\psi-\theta)\rho\big]\,.
\label{eq:F-P_mult}
\end{eqnarray}
This equation has an exact analytical expression for the stationary distribution $\rho_{s}(\theta)$, given by
\begin{equation}
\label{hypermises}
\rho_{s}(\theta)={\cal N}^{-1}\big[z+\sgn(\sigma)\sqrt{z^2-1}\cos(\psi-\theta)\big]^\nu\,,
\end{equation}
where $\sgn(\sigma)$ is the sign function. The normalization constant ${\cal N}$ and parameters $z$ and $\nu$ are
\begin{eqnarray}
\label{constn}
        &&{\cal N}=2\pi P_{\nu}^{0}(z)\,,\\
        &&z=(1-\sigma^{2}r^2)^{-1/2}\,,\\
	&&\nu=\frac{K}{D\sigma}-1=\frac{2}{T\sigma}-1\,,
\end{eqnarray}
where $P_{\nu}^{0}(z)$ is the associated Legendre function of zero order and $T=2D/K$ is the temperature of the system.

\section{Critical behavior}

We have already established the first law of thermodynamics of phase oscillators that takes the form
\begin{equation}
\label{Fhelm}
dF=-SdT-H_{s}dr\,,
\end{equation}
where $F=F(T,r)$ is the Helmholtz free energy, and $S$, $T$, and $r$ are entropy, temperature and order parameter, respectively. Moreover, we have defined a new quantity $H_{s}$ that plays the role of a synchronization field on the oscillator system. All these thermodynamics quantities have been previously obtained and studied in \cite{Pinto}. Here, we will use these results to analyze the critical behavior of the system.

\subsection{Order parameter}

From stationary density $\rho_{s}$, the order parameter $r$ can now be properly calculated
\begin{equation}
\label{order2}
r=\int_{0}^{2\pi}e^{i(\theta-\psi)}\rho_{s}(\theta)d\theta=\frac{\sgn{(\sigma)}}{1+\nu}\frac{P_{\nu}^{1}(z)}{P_{\nu}^{0}(z)}\,.
\end{equation}
Therefore, we can establish a critical region of the system for which $r\approx 0$, such that
\begin{equation}
z=(1-\sigma^{2}r^2)^{-1/2}\approx 1\,.
\end{equation}
We can now take the first terms of expanded  Eq. (\ref{order2}) for $z\approx 1$, which leads to
\begin{equation}
\label{r2}
r=\frac{\sqrt{2}\sgn(\sigma)\nu}{2}(z-1)^{1/2}-\frac{\sqrt{2}\sgn(\sigma)(\nu^2+\nu+1)\nu}{8}(z-1)^{3/2}\,,
\end{equation}
where for $r\approx 0$, we also assume
\begin{equation}
\label{z4}
z=(1-r^{2}\sigma^{2})^{-1/2}\approx 1+\frac{r^{2}\sigma^{2}}{2}+\frac{3r^{4}\sigma^{4}}{8}\,,
\end{equation}
and $(1+\frac{3}{4}r^{2}\sigma^{2})^{1/2}\approx 1+\frac{3}{8}r^{2}\sigma^{2}$ has been used.

\subsection{Entropy}

We have consistently shown that the free energy for a phase oscillators system with multiplicative noise is given by
\begin{equation}
F=-T\ln[2\pi P_{\nu}^{0}(z)]\,,
\label{F_geral}
\end{equation}
which allows us to obtain entropy $S$ by the definition Eq. (\ref{Fhelm}) as
\begin{equation}
\label{hyperentropy2}
S=-\left(\frac{\partial F}{\partial T}\right)_r = \Big(1-\nu\frac{\partial}{\partial\nu}\Big)\ln\big[2\pi P_{\nu}^{0}(z)\big]\,.
\end{equation}
We now analyze the entropy Eq. (\ref{hyperentropy2}) in the critical region by expanding the Legendre function for $z\approx 1$, which results in
\begin{equation}
P_{\nu}^{0}(z)\sim 1+\frac{\nu(\nu+1)(z-1)}{2}\,,
\end{equation}
where the critical entropy is given by
\begin{equation}
\label{criticals}
S=S_{max}-\frac{\nu^2\sigma^2}{4}r^2\,.
\end{equation}
Here $S_{max}=\ln(2\pi)$ is the maximum entropy of the system.

\subsection{Synchronization field}

The synchronization field is properly defined by the first law of thermodynamics Eq. (\ref{Fhelm}) as
\begin{equation}
\label{fullH}
H_{s}=-\left(\frac{\partial F}{\partial r}\right)_{T}= \frac{T(1+\nu)z^3\sigma^2r^{2}\sgn(\sigma)}{\sqrt{z^2-1}}\,.
\end{equation}
We have already shown that field the $H_{s}$ is associated with the noise effect of the system, i.e., the Gaussian white noise and multiplicative noise. Hence $H_{s}$ can be decomposed as
\begin{equation}
\label{Hbroken}
H_{s}= H_{0} + H_{\sigma}\,,
\end{equation}
where $H_{0}$ and $H_{\sigma}$ are the internal and external synchronization fields, respectively, defined as
\begin{eqnarray}
\label{HGmf}
&&H_{0}=\lim_{\sigma\rightarrow 0}H_{s}=2r=2\frac{I_{1}(2r/T)}{I_{0}(2r/T)}\\
\nonumber\\
&&H_{\sigma}=H_{s}(\sigma,r,T)-H_{0}\,,
\end{eqnarray}
in which $I_{0}(x)$ and $I_{1}(x)$ are the modified Bessel functions of first kind of order $0$ and $1$, respectively. Note that $H_{0}$ is associated to part of field $H_{s}$, which does not depend explicitly on $\sigma$, i.e., $H_{0}$ is the field related to the Gaussian white noise behavior. On the other hand, $H_{\sigma}$ corresponds to part of field $H_{s}$ which explicitly depends on $\sigma$. This is directly associated to the non null multiplicative noise.

Now expanding Eq. (\ref{fullH}) in the critical region $r\approx 0$, we obtain
\begin{equation}
\label{cfield}
H_{s}\sim \frac{1}{2}T\nu^{2}\sigma^{2}r+\frac{1}{2}T\nu\sigma^{2}r\,.
\end{equation}
Thus inserting $\nu=2/T\sigma-1$ in Eq. (\ref{cfield}) and retaining the terms in $r$, we obtain the critical synchronization field
\begin{equation}
\label{ccfield}
H_{s}=\frac{2r}{T}-\sigma r=H_{0}+H_{\sigma}\,.
\end{equation}
Note that $H_{0}=\lim_{\sigma\rightarrow 0}H_{s}=2r/T$ is the critical internal field and
\begin{equation}
\label{cHsig}
H_{\sigma}=-\sigma r\,,
\end{equation}
is the critical external field. Indeed, the field $H_\sigma$ depends explicitly on the noise coupling $\sigma$, as expected. The negative sign also shows that synchronization $r$ should decrease when $-\sigma$ increases in the critical region.

\subsection{Specific heat}

We can now define the critical specific heat at constant external field $H_{\sigma}$. First, we express critical entropy Eq. (\ref{criticals}) in terms of $H_{\sigma}$, Eq. (\ref{cHsig}), which gives
\begin{equation}
	S=S_{max}-\frac{r^2}{T^2}-\frac{H_{\sigma}}{T}r-\frac{H_{\sigma}^{2}}{4}.
\end{equation}
It follows that the critical specific heat at constant external field is given by
\begin{equation}
\label{sc11}
C_{H_{\sigma}}=T\left(\frac{\partial S}{\partial T} \right)_{H_{\sigma}}=\frac{2r^2}{T^2}-\frac{2r}{T}\frac{\partial r}{\partial T}+H_{\sigma}\Big(\frac{r}{T}-\frac{\partial r}{\partial T}\Big)\,.
\end{equation}
It is important to emphasize that we have achieved the main thermodynamic functions in the critical region for the phase oscillator system. Furthermore, the concept of critical synchronization field shown above is of crucial importance for determining the complete thermodynamics. Indeed, this reflects the precise determination of the main critical exponents for the model, as we will see below.

\subsection{Fluctuation-dissipation relation}

We can also examine the relation between order parameter fluctuation $\left<r^{2}\right>$ and its response function in the vicinity of the phase transition. We begin by considering the probability of fluctuation $w(r)$ of the order parameter which can be directly obtained as
\begin{equation}
\label{Sd}
w(r)\propto \exp{\left(\Delta S\right)}\,,
\end{equation}
where $\Delta S$ is the change in entropy in the fluctuation which can be expressed as $\Delta S=-W_{min}/T$ \cite{Landau}, such that $W_{min}$ is the minimum external work which must be performed on the system in order to reversibly produce this fluctuation. At the critical region, the minimum external work is then related to the external field $H_{\sigma}$, Eq. (\ref{cHsig}), according to
\begin{equation}
W_{min}=\int_{0}^{r}H_{\sigma}dr=-\frac{\sigma r^{2}}{2}\,.
\end{equation}
On the other hand, from the definition of susceptibility

\begin{equation}
\chi^{-1} =\left(\frac{\partial H_{\sigma}}{\partial r}\right)_{T}=-\sigma\,,
\end{equation}
which allows rewrite Eq. (\ref{Sd}) in the form
\begin{equation}
w(r)\propto \exp{\left(-\frac{r^2}{2\chi T}\right)}\,,
\end{equation}
which is the usual Gaussian distribution, whose mean square fluctuation $\left<r^{2}\right>$ takes the form
\begin{equation}
\left<r^{2}\right>=T_{c}\chi\,.
\label{fltdis}
\end{equation}
This is the fluctuation-dissipation relation for the oscillators system, relating its linear response $\chi$, due to the action of the external field $H_{\sigma}$, with the fluctuations of the order parameter around the equilibrium region $\left<r\right>=0$. Indeed, a similar relationship holds in magnetic systems for the magnetization fluctuation \cite{Falk}.

\section{Critical exponents}

We are now able to determine the main critical exponents for the phase oscillator system. First we must define, in the critical region, the quantity
\begin{equation}
\tau=T_c-T=1-T.
\end{equation}
with $T\approx T_{c}$ where $T_{c}=1$ is the critical temperature.

\subsection{Order parameter}

We start by determining the critical exponents of the order parameter $r$. Note that Eq. (\ref{r2}) can be written in terms of external field Eq. (\ref{cHsig}) as
\begin{equation}
\label{r20}
r^{2}\Lambda=1-\frac{1}{T}+\frac{\sigma}{2}=1-\frac{1}{T}-\frac{H_{\sigma}}{2r}\,,
\end{equation}
where
\begin{equation}
\label{r21}
\Lambda=\frac{\nu\sigma^3}{16}\big(3-\nu^2-\nu-1\big)=\frac{H_{\sigma}^{3}}{8r^3}+\frac{H_{\sigma}^{2}}{8rT}-\frac{H_{\sigma}}{2rT^2}-\frac{1}{2T^3}\,.
\end{equation}

We should now impose that $T\rightarrow T_{c}=1$ for null field $H_{\sigma}=0$, such that Eq.(\ref{r21}) goes to $\Lambda=-\frac{1}{2}$ and Eq. (\ref{r20}) converges to
\begin{equation}
\label{ordbe}
r=\lim_{\tau\rightarrow 0}\Big[\Lambda^{-1}\big(1-\frac{1}{T}\big)\Big]^{1/2}\propto (1-T)^{1/2}\propto \tau^{\beta},
\end{equation}
where $\beta=\frac{1}{2}$ is the critical exponent for the order parameter $r$, as has already been obtained for the Kuramoto model \cite{Basnarkov1,Crawford}. Here, however, we have presented a rigorous demonstration through the synchronization field concept. Furthermore, note also that along the isotherm $T=T_{c}=1$,  Eq. (\ref{r20}) leads to
\begin{equation}
\label{rr33}
r^{3}\Lambda=-\frac{1}{2}H_{\sigma}\,,
\end{equation}
and taking $\tau=0$ and $H_{\sigma} \rightarrow 0$ (with $\Lambda=-1/2$), we get the relation between $r$ and external field $H_{\sigma}$ as
\begin{equation}
\label{orddel}
r=-\lim_{H_{\sigma}\rightarrow 0}\frac{1}{(2\Lambda)^{1/3}} H_{\sigma}^{1/3}\propto H_{\sigma}^{1/\delta},
\end{equation}
where we obtain the critical exponent $\delta=3$ for the system.

\subsection{Specific heat}

In order to obtain the critical exponent for the specific heat, we start from Eq.(\ref{sc11}) for null field $H_{\sigma}=0$ and $\tau\rightarrow0$, which leads to
\begin{equation}
 \label{CH}
 C_{H_{\sigma}=0}=\lim_{\tau\rightarrow 0}\left[\frac{2r^2}{T^2}-\frac{2r}{T}\frac{\partial r}{\partial T}\right]=\left(-2r\frac{\partial r}{\partial T}\right)_{r= \tau^{\frac{1}{2}}}\propto \tau^{\frac{1}{2}}\tau^{-\frac{1}{2}}\propto \tau^{\alpha}\,,
\end{equation}
where $\alpha=0$ is the critical exponent for the specific heat.

\subsection{Susceptibility}

Now, from the definition of inverse susceptibility $\chi^{-1}$ and taking Eq. (\ref{rr33}) for the null field $H_{\sigma}=0$, $\Lambda=-1/2$, and $\tau\rightarrow 0$, we obtain
\begin{equation}
\label{sus}
\chi^{-1} =\left( \frac{\partial H_{\sigma} }{ \partial r} \right)_{H_{\sigma}=0}\propto r^{2}\propto \tau\,,
\end{equation}
where we use $r\propto \tau^{1/2}$. This strictly shows that the susceptibility is divergent, i.e.,
\begin{equation}
\label{sus2}
\chi\propto \tau^{-\gamma},
\end{equation}
with critical exponent $\gamma=1$, in precise accordance with the mean field theory. According to Eq.~(\ref{sus2}), the mean square fluctuation of the order parameter Eq.~(\ref{fltdis}) increases as $1/\tau$ when $T\rightarrow T_{c}$.

These results conclude the study of the four main critical exponents of the generalized Kuramoto model for which we have also shown the fluctuation-dissipation relation. It is important to note that the critical exponents $\beta=\frac{1}{2}$, $\alpha=0$, $\delta=3$, and $\gamma=1$ are the mean field exponents which satisfy $\alpha +2\beta+\gamma=2$ and $\gamma=\beta(\delta-1)$, which are the Rushbrooke and Widom scaling laws, respectively.

\section{Conclusions}

In this article we have systematically studied the critical behavior of phase oscillators with multiplicative noise from a thermodynamic equilibrium approach. We derived the set of the four main mean-field critical exponents $\alpha=0$, $\beta=1/2$, $\gamma=1$ and $\delta=3$ for the system which obey the universal scale laws of Rushbrooke and Widom. Indeed, this is the first time that all of these exponents have been presented for phase oscillator systems. The critical behavior associated with phase oscillators may appear in many physical systems, such as biomolecular networks and neural systems \cite{Plentz,Luonan,Kromer}, in particular in neuronal avalanches phenomenon \cite{Yu}, where synchronization transition is present. Furthermore, susceptibility and specific heat as presented in this article can play important roles in the description of the thermodynamics of these physical systems.

\section{Acknowledgments}

 We acknowledge the support of the Conselho Nacional de Desenvolvimento Cient\'{i}fico e Tecnol\'{o}gico (CNPq) Brazil, the Coordena\c{c}\~{a}o de Aperfei\c{c}oamento de Pessoal de N\'{i}vel Superior (CAPES) Brazil, and the Funda\c{c}\~{a}o de Apoio \`{a} Pesquisa do Distrito Federal (FAP-DF), Brazil.


\section*{References}
\begin{thebibliography}{99}


\bibitem{Kuramoto1} Y. Kuramoto, Chemical Oscillations, Waves, and Turbulence (Springer-Verlag, Berlin, 1984).

\bibitem{Basnarkov1} L. Basnarkov and V. Urumov, Phys. Rev. E {\bf 76}, 057201 (2007).

\bibitem{Basnarkov2} L. Basnarkov and V. Urumov, Phys. Rev. E {\bf 78}, 011113 (2008).

\bibitem{Ewa} M. Bier, B. Lisowski, and E. Gudowska-Nowak, Phys. Rev. E {\bf 93}, 012143 (2016).

\bibitem{Daido1} H. Daido, Phys. Rev. Lett. {\bf 73}, 760 (1994).

\bibitem{Crawford} J. D. Crawford, Phys. Rev. Lett. {\bf 74}, 4341(1995).

\bibitem{Daido2} H. Daido, Phys. Rev. E {\bf 91}, 012925 (2015).

\bibitem{Yoon} S. Yoon, et al. Phys. Rev. E {\bf 91}, 032814 (2015).

\bibitem{Hong} H. Hong, et al. Phys. Rev. E {\bf 92}, 022122 (2015).

\bibitem{Pinto} P. D. Pinto, F.A. Oliveira, and A.L.A. Penna, Phys. Rev. E {\bf 93}, 052220 (2016).

\bibitem{Lai} Y. M. Lai and M. A. Porter, Phys. Rev. E {\bf 88}, 012905 (2013).

\bibitem{Yoshimura} K. Yoshimura and K. Arai, Phys. Rev. Lett. {\bf 101}, 154101 (2008).

\bibitem{Teramae2} J. Teramae, H. Nakao, and G.B. Ermentrout, Phys. Rev. Lett. {\bf 102}, 194102 (2009).

\bibitem{Erdelyi} A. Erdelyi et al., {\it Higher Transcendental Functions} Vol. I (McGraw-Hill, New York, 1953).

\bibitem{Landau} L. D. Landau and E. M. Lifshitz, {\it Statistical Physics} Vol. 5 (Oxford: Pergamon Press, 1980).

\bibitem{Falk} H. Falk and L. W. Bruch, Phys. Rev. {\bf 180}, 442 (1969).

\bibitem{Plentz} D. Plenz, E. Niebur, and H. G. Schuster, {\it Criticality in Neural Systems} (Wiley-VCH Verlag: Weinheim, Germany, 2014).

\bibitem{Luonan} L. Chen et al., {\it Modelling Biomolecular Networks in Cells} (Springer: London, 2010).

\bibitem{Kromer} J.A. Kromer, L. Schimansky-Geier, and A.B. Neiman, Phys. Rev. E {\bf 93}, 042406 (2016).

\bibitem{Yu} S. Yu, H. Yang, O. Shriki, and D. Plenz, Front. Syst. Neurosci. {\bf 7}, 42 (2013).





%
%
%
%
%
%
%
%
%
%
%
%
%
%
%
%
%
%
%
%
%
%
%
%
%
%
%
%
%
%
%
%
%
%
%
%
%
%
%
%
%
%
%
%
%
%
%
%
%
%









\end{thebibliography}
\end{document}